\documentclass[letterpaper,twocolumn,english,showpacs,prl]{revtex4}
\usepackage{times}
\usepackage[T1]{fontenc}
\usepackage[latin1]{inputenc}
\usepackage{amsmath}
\usepackage{graphicx}
\usepackage{amssymb}

\makeatletter
\usepackage{babel}
\makeatother
\begin{document}

\title{The Intrinsic Time Scale of Transient Neuronal Responses}

\author{Björn Naundorf$^{1}$, Theo Geisel$^{1,2}$, and Fred Wolf$^{1,2}$}

\affiliation{$^{1}$Max-Planck-Institut für Strömungsforschung and Fakultät für
Physik, Universität Göttingen, 37073 Göttingen, Germany%
\footnote{Present and permanent address. %
}}

\affiliation{$^{2}$Kavli Institute for Theoretical Physics, University of California
Santa Barbara, CA 93106-4030}

\begin{abstract}
In a generic neuron model, we present the linear response theory for
the firing rate in response to both time dependent input currents
and noise amplitudes. In both cases the signal transmission is strongly
attenuated for frequencies above the stationary firing rate. For high
frequencies both the mean input and the noise transmission function
decay as $\omega^{-2}$, independent of model details. Our results
indicate that previously suggested mechanisms for near instantaneous
transmission of information are not consistent with the spike generation
mechanism of real neurons. 
\end{abstract}


\maketitle
In the brain information is processed through a hierarchy of neural
layers. Based on the observation that human subjects can successfully
distinguish complex visual stimuli in only a few hundred milliseconds
\cite{Thorpe}, it has been argued that the processing time of individual
layers must be very small. Recently two different mechanisms have
been proposed for the practically instantaneous relaying of information
at the single neuron level. They utilize the fact that in vivo cortical
neurons exhibit substantial subthreshold fluctuations of their membrane
potentials, due to the large number of presynaptic neurons \cite{DestexhePare}.
The first mechanism assumes that information is transferred by a time
dependent noise amplitude \cite{SigmaModulation1,SigmaModulation2},
the second assumes information transfer via a modulation of the mean
input in the presence of temporally correlated background noise \cite{Gerstner,Fourcaud}.
These studies, which are based on a seminal work by Knight \cite{Knight},
have been conducted on integrate-and-fire models. These models, however,
are highly idealized models of cortical neurons. Their main ingredient
is a fixed voltage threshold. Each time the membrane potential reaches
this threshold, a spike is said to be emitted. Real neurons, however,
as well as biophysically realistic conductance based neuron models
do not have a fixed voltage threshold and are rather comparable to
excitable systems. This naturally raises the question whether the
response properties found in simple threshold models are preserved
if one considers models with a more realistic spike generating mechanisms.
In this letter we address this question by a combination of analytical
and numerical techniques considering a generic Type-I neuron model,
the $\theta$-Neuron \cite{GutkinErmentrout}. We show that in a realistic
regime the response to a transient stimulus, both in the mean input
as well as in the noise amplitude is \emph{not} instantaneous as predicted
by the integrate-and-fire models. Moreover, we show that the maximum
transmission frequency is approximately determined by the stationary
firing rate. Thus our results suggest that the $\theta$-Neuron reproduces
the dynamical behavior of cortical neurons faithfully, while still
being tractable analytically.

Neurons have been divided into several classes based on the kind of
excitability they exhibit. In our study we focus on Type-I excitability
which can be encountered in many neurons withing the cortex \cite{HodgkinHuxley}.
In models of Type-I neurons, repetitive firing typically emerges via
a saddle-node bifurcation. The normal form of this bifurcation is
equivalent to a phase oscillator, the $\theta$-Neuron \cite{GutkinErmentrout}.
It has been shown that it has the same mechanism of excitability and
super-threshold behavior as cortical neurons \cite{HanselMato}. Its
dynamics is:\[
\tau\dot{\theta}=(1-\cos\theta)+I(t)(1+\cos\theta),\]
where $\tau$ is the time constant and $I(t)$ represents the total
synaptic input to the neuron. Each time the oscillator crosses the
point $\theta=\pi$ a spike is said to be {}``fired''. For constant
inputs this model exhibits a periodic firing regime for $I>I_{c}=0$
with firing rate $\nu=\sqrt{I}/\left(\pi\tau\right)$, and an excitable
regime for $I<0$. We analyze the response of an ensemble of such
neurons, which is given by the ensemble averaged firing rate $\nu(t)$.
The input current $I(t)$ is decomposed,\[
I(t)=I_{0}+\sigma\sqrt{\tau}z(t),\]
 into a mean current $I_{0}$ and a noise term which reflects the
fluctuations induced by the synaptic inputs \cite{Tuckwell}. The
noise is modeled by an Ornstein-Uhlenbeck process with a correlation
time $\tau_{c}$:\[
\tau_{c}\frac{dz}{dt}=-z+\eta(t),\]
where $\eta(t)$ is Gaussian white noise. In the limit $\tau_{c}\to0$,
$z(t)$ becomes white noise. In the following, we have chosen the
time constant $\tau=0.25\textrm{ms}$ which results in a spike duration
of about $1\textrm{ms}$ as found in real neurons. $I_{0}$ and $\sigma$
have been chosen to give realistic $\theta$-correlation times of
approximately $10\textrm{ms}$ and firing rates in the range $1-20\textrm{Hz}$.

The state of an ensemble of such neurons is described by a probability
density function $P(\theta,z,t)$. Its dynamics is determined by the
Fokker-Planck equation:\begin{equation}
\partial_{t}P=\hat{L}P,\label{eq:FokkerPlanck}\end{equation}
with the operator,\begin{eqnarray*}
\hat{L} & = & -\tau^{-1}\partial_{\theta}\left(\left(1-\cos\theta\right)+\left(I_{0}+\sigma\tau^{1/2}z\right)\left(1+\cos\theta\right)\right)\\
 &  & +\tau_{c}^{-1}\partial_{z}z+\frac{1}{2}\tau_{c}^{-2}\partial_{z}^{2},\end{eqnarray*}
 periodic boundary conditions in the $\theta$-direction and natural
boundary conditions in the $z$-direction. The firing rate $\nu(t)$
is identical to the total probability current through the line $\theta=\pi$:\[
\nu(t)=2\tau^{-1}\int_{-\infty}^{\infty}P(\pi,z,t)\, dz.\]

In the limit of temporally uncorrelated input, $\tau_{c}\to0$, the
rate can be calculated analytically \cite{Risken,LindnerBulsara}
which gives:\[
\nu_{\textrm{WN}}^{-1}=\frac{4\tau\sqrt{\pi}}{\sigma}\int_{0}^{\infty}dy\,\exp\left\{ -\frac{4}{\sigma^{2}}\left(\frac{y^{6}}{3}+I_{0}y^{2}\right)\right\} \]
To investigate how the neuron responds to time dependent synaptic
inputs it is important to consider both the response to a time dependent
mean input current $I=I_{0}+\epsilon e^{i\omega t}+\sqrt{\tau}\sigma z$
and a time dependent noise amplitude $I=I_{0}+\sqrt{\tau}\left(\sigma+\epsilon e^{i\omega t}\right)z$.
Inserting the modulated input current into Eq.~(\ref{eq:FokkerPlanck})
and expanding $P(\theta,z,t)=P_{0}(\theta,z)+\epsilon\tilde{P}_{\omega}(\theta,z)e^{i\omega t}+\cdots$
gives in linear order in $\epsilon$ \cite{Risken}:\begin{equation}
i\omega\tilde{P}_{\omega}(\theta,z)=\hat{L}\tilde{P}_{\omega}-f(\theta,z),\label{eq:LinearResponseDeq}\end{equation}
with $f(\theta,z)=-\tau^{\textrm{-1}}\partial_{\theta}(1+\cos\theta)P_{0}(\theta,z)$
for modulated input currents and $f(\theta,z)=-\tau^{-1/2}\partial_{\theta}\left(1+\cos\theta\right)zP_{0}(\theta,z)$
for modulated noise amplitudes. The formal solution of eq.~(\ref{eq:LinearResponseDeq})
is:\begin{equation}
\tilde{P}_{\omega}(\theta,z)=e^{i\omega t}\int_{-\infty}^{t}e^{(t-t')\hat{L}}f(\theta,z)\, e^{i\omega t'}\, dt'.\label{eq:LinearResponseIntegral}\end{equation}
 This integral can be solved in terms of eigenfunctions $P_{k}(\theta,z)$
of the operator $\hat{L}$:\begin{equation}
\lambda_{k}P_{k}(\theta,z)=\hat{L}P_{k}(\theta,z),\label{eq:EigenvalueOperatorProblem}\end{equation}
with the associated eigenvalues $\lambda_{k}$. Because of the applied
boundary conditions the spectrum $\left\{ \lambda_{k}\right\} $ is
discrete, and since detailed balance is not fulfilled, eigenvalues
$\lambda_{k}$ and the corresponding eigenfunctions $P_{k}(\theta,z)$
are either real or form complex conjugate pairs. Moreover, the eigenfunctions
are typically not orthogonal. However, an orthonormal set of basis
functions $\left\{ \left|\phi_{k}\right\rangle \right\} $ can be
constructed from them by Gram-Schmidt orthogonalization (in Dirac
notation) with expansion coefficients $a_{l}^{k}$:\[
\left|\phi_{k}\right\rangle (\theta,z)=\sum_{l=1}^{k}a_{l}^{k}P_{l}(\theta,z).\]
Inserting $\sum_{k}\left|\phi_{k}\right\rangle \left\langle \phi_{k}\right|=\mathbf{1}$
into Eq.~(\ref{eq:LinearResponseIntegral}) yields,\begin{eqnarray*}
\tilde{P}_{\omega}(\theta,z) & = & e^{-i\omega t}\sum_{k,l\leq k}b_{k}a_{l}^{k}\int_{-\infty}^{t}e^{i\omega t'+(t-t')\hat{L}}P_{l}(\theta,z)\\
 & = & \sum_{k,l\leq k}\frac{a_{l}^{k}b_{k}}{i\omega-\lambda_{l}}P_{l}(\theta,z),\end{eqnarray*}
with $b_{k}=\left\langle \phi_{k}|f\right\rangle $. The rate response
is then given by:\begin{eqnarray*}
\nu(t) & = & 2\tau^{-1}\int_{-\infty}^{\infty}\left(P_{\textrm{st}}(\pi,z)+\tilde{P}_{\omega}(\pi,z)e^{i\omega t}\right)\, dz\\
 & = & \nu_{0}+\nu_{1}(\omega)e^{i(\omega t+\phi(\omega))}.\end{eqnarray*}
The limit $\omega\to\infty$ can be treated analytically. Since $f(\theta,z)$
vanishes at $\theta=\pi$, the modulus of $\tilde{P}$ at this point
has to be proportional to $\omega^{-2}$:\[
\left(i\omega-\hat{L}\right)\tilde{P}_{\omega}(\pi,z)e^{-i\omega t}=-f(\pi,z)=0.\]
 An expansion in $\omega^{-1}$ reveals that the relative phase of
$\nu_{1}(\omega)$ is $-\pi$ in this limit. We would like to stress
that this decay is universal and does not depend on model details.
It is only due to the insensitivity to external inputs at the point
where a spike is emitted, i.e.~$f(\pi,z)=0$.

\begin{figure}[htbp]
\includegraphics[%
  width=1.0\columnwidth,
  keepaspectratio]{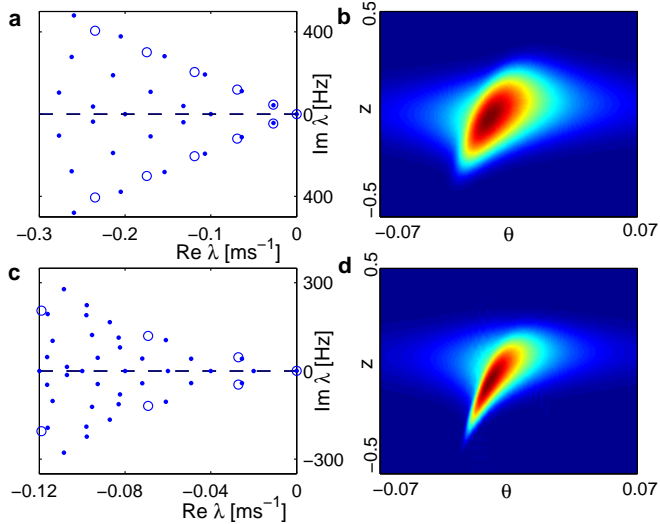}

\caption{\label{cap:Spectrum}Spectrum of operator $\hat{L}$ (a,c) and stationary
density (b,d) for $\tau_{c}=10\textrm{ms}$, $\nu_{0}=10\textrm{Hz}$
(a,b) and $\tau_{c}=50\textrm{ms}$, $\nu_{0}=8\textrm{Hz}$ (c,d)
\cite{NoteOnNum}. The open circles denote the result in the limit
$\tau_{c}\to0$. In all cases $\sigma=10^{-3}$, $I_{0}=0$. For increasing
$\tau_{c}$ the tips of the wedges move closer to the imaginary axis.}
\end{figure}

The eigenvalues and eigenfunctions of the time-independent operator
$\hat{L}$ determine $\nu_{1}(\omega)$ completely. We computed them
using a matrix representation of Eq.~(\ref{eq:EigenvalueOperatorProblem})
obtained by expanding $P_{k}(\theta,z)$ into a complete set of orthonormal
functions:\begin{eqnarray*}
P_{k}(\theta,z) & = & \sum_{m=0}^{\infty}\sum_{n=-\infty}^{\infty}a_{n,m}\left(2^{n+1}\sqrt{\pi/2\tau_{c}}\, m!\right)^{-1/2}\\
 &  & \qquad\qquad e^{in\theta}H_{m}(\sqrt{2\tau_{c}}z)e^{-z^{2}\tau_{c}}.\end{eqnarray*}
Here $H_{m}(z)$ are the Hermite polynomials \cite{AbramowitzStegun}.
Inserting this into Eq.~(\ref{eq:EigenvalueOperatorProblem}), multiplying
from left with $\left(2^{n'+1}\sqrt{\pi/2\tau_{c}}\, m'!\right)^{-1/2}e^{in'\theta}H_{m'}(\sqrt{2\tau_{c}}z)e^{-z^{2}\tau_{c}}$
and integrating over the entire domain leads to the following eigenvalue
problem:\begin{equation}
\lambda a_{n,m}=\sum_{n',m'}L_{n,m;n',m'}a_{n',m'}.\label{eq:MatrixEigenProblem}\end{equation}
Because the Fokker-Planck operator has only two Fourier components
in the $\theta$-direction and is polynomial in the $z$-direction,
$\mathbf{L}$ is sparse with:\begin{eqnarray*}
L_{n,m;n',m'} & = & -i\tau^{-1}(1+I_{0})n-\tau_{c}^{-1}m\\
L_{n,m;n'\pm1,m'} & = & (2\tau)^{-1}i(1-I_{0})n\\
L_{n,m;n',m'-1} & = & -in\sigma\left(4\tau\tau_{c}\right)^{-1/2}\left(m+1\right)\\
L_{n,m;n',m'+1} & = & -in\sigma\left(4\tau\tau_{c}\right)^{-1/2}m\\
L_{n,m;n'\pm1,m'\pm1} & = & -in\sigma\left(16\tau\tau_{c}\right)^{-1/2}(m+1)\\
L_{n,m;n',m'-2} & = & \left(\sqrt{2\tau\tau_{c}}\right)^{-1}\sqrt{(m+1)(m+2)}\end{eqnarray*}
 We solve (\ref{eq:MatrixEigenProblem}) numerically using the Arnoldi-method
\cite{Arnoldi}, a high performance iterative algorithm. Figure \ref{cap:Spectrum}
shows the spectrum and the stationary density for two different correlation
times $\tau_{c}$ together with the spectrum in the white noise limit.
The eigenvalues are arranged in a series of wedges. The tip of each
wedge is located at integer multiples of $\tau_{c}^{-1}$. The different
wedges account for the decay of excitations in the $z$-direction,
whereas the eigenvalues within a wedge account for the decay (real
part) and oscillation (imaginary part) of excitations in the $\theta$-direction.
In the limit $\tau_{c}\to0$ the spectrum exhibits only one wedge
(open circles). For $\tau_{c}=10\textrm{ms}$ (upper plot), the interaction
between eigenvalues leads to deviations from the white noise limit
only for strongly damped modes. The deviations become more pronounced
only for large values of $\tau_{c}=50\textrm{ms}$ (lower plot). In
both regimes the stationary density strongly deviates from a separable
density.%

\begin{figure}[htbp]
\includegraphics[%
  width=1.0\columnwidth,
  keepaspectratio]{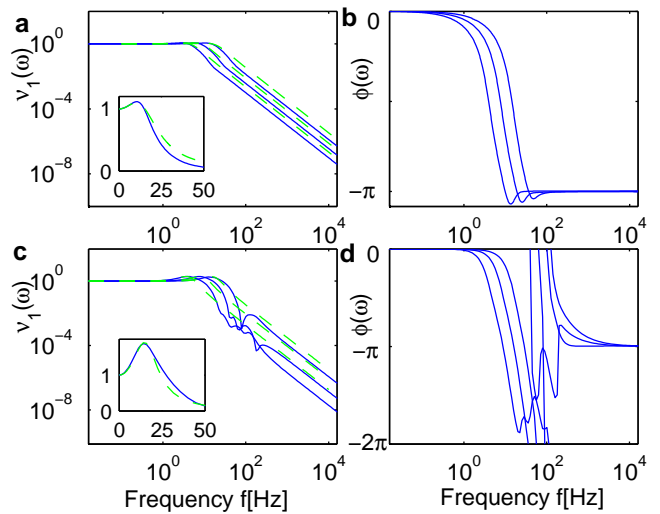}

\caption{\label{cap:LinearResponse}Response amplitude (a,c) and phase (b,d)
for a modulation in the input current (a,b) and in the noise amplitude
(c,d). The different curves correspond to different values of $\nu_{0}$
($5,10,20\textrm{Hz}$) and $\tau_{c}=10\textrm{ms}$. In the case
of a modulated input current the amplitude shows a maximum at approximately
the stationary firing rate and decays then rapidly proportional to
$\omega^{-2}$ (inset for $\nu_{0}=20\textrm{Hz}$, dashed: Lorentzian
approximation). The relative phase lag drops from zero to $-\pi$
and shows a small dip at the resonance frequency. For a modulated
noise amplitude there are more resonances at higher frequencies, but
these are strongly damped.}
\end{figure}
Examples of the response amplitude $\nu_{1}(\omega)$ and the phase
$\phi(\omega)$ are depicted for different values of $\nu_{0}$ in
Fig.~\ref{cap:LinearResponse} for both mean input and noise stimulation.
In the case of mean input modulation, the linear response amplitude
exhibits a resonance maximum at approximately the frequency of its
stationary firing rate and then decays rapidly to zero. The response
phase starts at zero and then drops to $-\pi$. For a modulation in
the noise amplitude the behavior is similar, except for additional
resonances at higher frequencies, which are, however, strongly damped. 

Whereas in general all eigenfunctions and eigenvalues contribute to
the rate response, the relatively simple Lorentz-like shape of the
response functions suggests that one frequency effectively dominates
the neurons' behavior. We observed that this cut-off frequency is
given by the imaginary part of the first excited eigenvalue of the
first wedge for the current response, and by the corresponding eigenvalue
of the second wedge for the noise response. This is demonstrated by
the green dashed lines in Fig.~\ref{cap:LinearResponse}. These are
the sum of two Lorentzians with their maxima at the positive and negative
frequency given by the imaginary part of the second eigenvalue in
the corresponding wedge. The reason for this is apparent from for
the decomposition of $f(\theta,z)$ into eigenfunctions of the Fokker-Planck
operator $\hat{L}$: \[
f(\theta,z)=\sum_{l}\alpha_{l}P_{l}(\theta,z).\]
 It has no contribution from the stationary density, since $\int f(\theta,z)\, d\theta=0$,
and for mean input modulations has a maximum contribution for the
eigenfunctions corresponding to second eigenvalue in the first wedge.
For noise modulations it has a maximum contribution for the eigenfunction
corresponding to the second eigenvalue in the second wedge. The imaginary
parts of the second eigenvalues in the first two wedges are, however,
almost identical for the parameters used. 

\begin{figure}[htbp]
\includegraphics{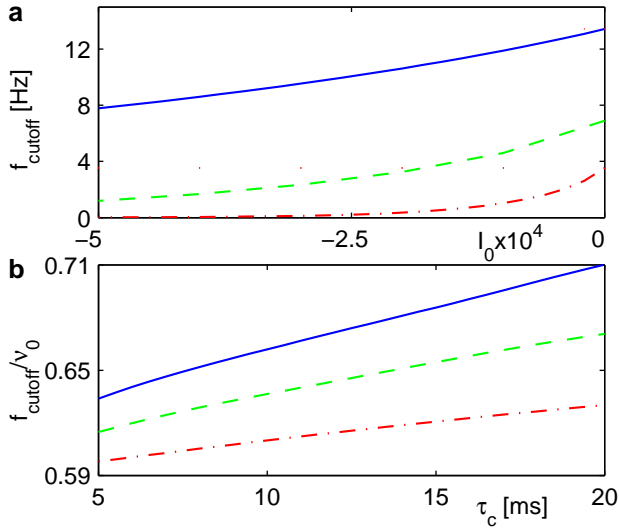}

\caption{\label{cap: Cutoff_frequency}Dependency of the cut-off frequency
for different values of the stationary firing rate $\nu_{0}$ ($20\textrm{Hz}$(solid),
$10\textrm{Hz}$(dashed), $5\textrm{Hz}$(dash-dotted)) as a function
of $I_{0}$ (a) and as a function of the correlation time $\tau_{c}$
compared to the stationary firing rate (b). For increasing values
of $\tau_{c}$ and $I_{0}$ the cut-off frequency increases, but always
stays below $\nu_{0}$. }
\end{figure}
The cut-off frequency for subthreshold mean inputs $I_{0}<0$ is depicted
in Fig.~\ref{cap: Cutoff_frequency} together with the dependence
on the noise correlation time $\tau_{c}$. The cut-off frequency increases
for increasing values of $I_{0}$ but always stays below the stationary
rate $\nu_{0}$. Increasing the noise correlation time at $I_{0}=0$
shifts the cut-off frequency to slightly larger values compared to
$\nu_{0}$. 

Thus our analysis demonstrates that in a realistic regime responses
much faster than $\nu_{0}^{-1}$ are strongly damped, since the transmission
function decays as $\omega^{-2}$. We would like to conclude this
letter with a comparison of the dynamic behavior of the $\theta$-neuron
with a biophysical realistic conductance based neuron \cite{WangBuszaki}
as well as with the classical leaky integrate-and-fire (LIF) model.
\begin{figure}[htbp]
\includegraphics{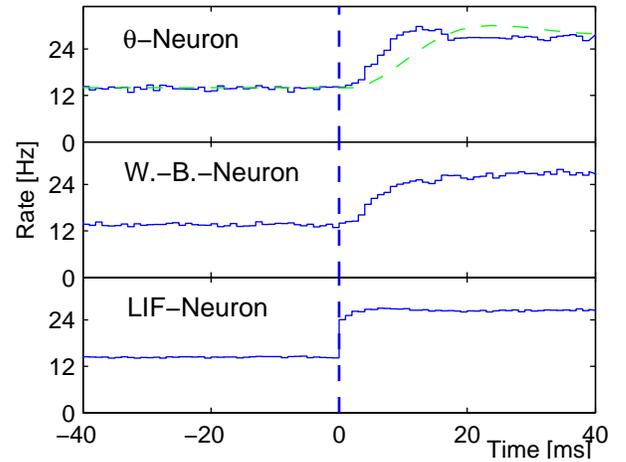}

\caption{\label{cap:CompThetaWBLIF}Comparison between the response of the
$\theta$-neuron (a), a conductance based model (b) and the LIF model
(c) to a voltage step. The model neurons are approximately at the
onset of repetitive firing, comparable to neurons in vivo and receive
an additional correlated input current with $\tau_{c}=10\textrm{ms}$.
The correlation time of the membrane potential is approximately $16\textrm{ms}$
for the $\theta$-neuron and the Wang-Buszáki neuron and approximately
$10\textrm{ms}$ for the LIF neuron (parameters as in \cite{Fourcaud}).
While the LIF neuron responds practically instantaneously, the response
time of the $\theta$-neuron and the conductance based model is about
$10\textrm{ms}$. (dashed: Linear response result).}
\end{figure}
Figure \ref{cap:CompThetaWBLIF} shows the step response of the three
models for an identical current correlation, identical initial and
final firing rates and similar membrane potential correlation times.
The dynamics of the $\theta$-neuron and the conductance based model
agree well, their response time is about $10\textrm{ms}$. The dashed
line shows the linear response result, which predicts a slower response
but is of the same order of magnitude. The LIF model on the other
hand responds practically instantaneously. This is due to the fact
that in the LIF model high input frequencies are not substantially
damped, i.e.~the response amplitude does not decay for large frequencies.
This is impossible in the $\theta$-neuron in which the response amplitude
always decays as $\omega^{-2}$. We would like to stress that this
decay is a universal property due to the insensitivity to input currents
at the point where a spike is fired and is independent of model details.
Thus, although one observes that the response times of both, the $\theta$-neuron
and the LIF model decrease with increasing $\tau_{c}$, the mathematical
origin and nature of this dependence is very different. Whereas in
the $\theta$-neuron it is a consequence of the dependence of the
eigenvalues on $\tau_{c}$, in integrate-and-fire models they result
from the voltage threshold.

In conclusion, we presented the linear response theory for the firing
rate of the $\theta$-neuron in response to both time dependent input
currents and time dependent noise amplitudes. For an effective numerical
treatment we derived a sparse matrix representation of the Fokker-Planck
operator. Using the eigenvalues and eigenfunctions of this operator,
we showed that the transmission amplitude is in both cases strongly
damped for frequencies above a cut-off frequency. In a wide range
of parameters this cut-off frequency is always below the mean firing
rate. We showed that the response behavior agrees well with the dynamics
of a conductance based model neuron and is different from the behavior
of the LIF model. Our results indicate that the $\theta$-neuron,
although simple, captures well the dynamical properties of real neurons.
They also reveal that previously proposed mechanisms of practically
instantaneous transmission of information are incompatible with the
spike generating mechanism of real neurons. Our findings suggest that
the $\theta$-neuron reproduces the dynamical behavior of cortical
neurons with Type-I excitability faithfully, while still being tractable
analytically.

We acknowledge useful discussion with M.~Bethge, P.~Latham, A.~Morrison,
K.~Pawelzik, M.~Timme and C.~v.~Vreeswijk. This research was supported
in part by the National Science Foundation under Grant No. PHY99-07949.


\begin{thebibliography}{10}
\bibitem{NoteOnNum}The basis for the numerical diagonalization has been truncated to
$n=-N\ldots N$ in the $\theta$- and $k=0\ldots K$ in the $z$-direction
with $N=6000$ and $K=40$. This results in a matrix of size $\left(4.8\cdot10^{5}\right)^{2}$
for which we computed up to the first $200$ eigenvalues and eigenfunctions.
\end{thebibliography}
\end{document}